\documentclass[12pt]{article}
\usepackage{amsmath,amssymb,graphicx,amsthm}
\usepackage{hyperref}
\numberwithin{equation}{section}

\begin{document}

\newcommand{\der}{\partial}
\newcommand{\e}{\mathrm{e}}
\newcommand{\ii}{\mathrm{i}}
\newcommand{\dd}{\mathrm{d}}
\newcommand{\vb}[1]{\mathbf{#1}}
\newcommand{\abs}[1]{\lvert#1\rvert}
\newcommand{\bx}{\mathbf x}
\newcommand{\by}{\mathbf y}
\newcommand{\bp}{\mathbf p}
\newcommand{\bq}{\mathbf q}
\newcommand{\br}{\mathbf r}
\newcommand{\bk}{\mathbf k}
\newcommand{\vf}{\varphi}
\newcommand{\eps}{\varepsilon}
\newcommand{\nr}{\eta_\rho}
\newcommand{\RR}{\mathbb R}
\newcommand{\NN}{\mathbb N}
\newcommand{\mM}{\mathcal{M}}
\newcommand{\mL}{\mathcal{L}}
\newcommand{\tend}{\rightarrow}
\newcommand{\then}{\Rightarrow}
\newcommand{\Li}{\mathrm{Li}}
\newcommand{\Tr}{\mathrm{Tr}}
\newcommand{\tttbr}{\langle\theta_{tt}\rangle_{\beta,\rho}}
\newcommand{\tttbry}{\langle\theta_{tt}(y)\rangle_{\beta,\rho}}
\newcommand{\tttbrr}{\langle\theta_{tt}(r)\rangle_{\beta,\rho}}

\newtheorem{theo}{Theorem}[section]
\newtheorem{coro}[theo]{Corollary}
\newtheorem{prop}[theo]{Proposition}
\newtheorem{defi}[theo]{Definition}
\newtheorem{conj}[theo]{Conjecture}
\newtheorem{lem}[theo]{Lemma}
\newcommand{\prf}{\underline{\it Proof.}\ }
\newcommand{\finprf}{\null \hfill {\rule{5pt}{5pt}}\\[2.1ex]\indent}

\pagestyle{empty} \rightline{September 2005}

\vfill

\begin{center}
{\Large\bf Aspects of Finite Temperature Quantum Field Theory in a
Black Hole Background}
\\[2.1em]

\bigskip

{\large Giuseppe Milanesi $^{a,}$\footnote{milanesi@sissa.it}
and Mihail Mintchev $^{b,}$\footnote{mintchev@df.unipi.it}}\\

\null

\noindent {\it $^a$ Scuola Internazionale Superiore di Studi
Avanzati, Via Beirut 2-4, 34014 Trieste, Italy and INFN, Sezione
di Trieste}
\\[2.1ex]
{\it $^b$ INFN and Dipartimento di Fisica, Universit\'a di Pisa,
Largo Pontecorvo 3, 56127 Pisa, Italy} \vfill

\end{center}

\begin{abstract}
We quantize a scalar field at finite temperature $T$ in the
background of a classical black hole, adopting 't Hooft's ``brick
wall'' model with generic mixed boundary conditions at the brick
wall boundary. We first focus on the exactly solvable case of two
dimensional space-time. As expected, the energy density is
integrable in the limit of vanishing brick wall thickness only for
$T=T_H$ - the Hawking temperature. Consistently with the most
general stress energy tensor allowed in this background, the
energy density shows a surface contribution localized on the
horizon. We point out that the usual divergences occurring in the
entropy of the thermal atmosphere are due to the assumption that
the third law of thermodynamics holds for the quantum field in the
black hole background. Such divergences can be avoided if we
abandon this assumption. The entropy density also has a surface
term localized on the horizon, which is open to various
interpretations. The extension of these results to higher
space-time dimensions is briefly discussed.
\end{abstract}

\vfill \rightline{SISSA 63/2005/EP}
 \rightline{IFUP-TH 19/2005}
\rightline{\tt hep-th/0509080}
\newpage
\pagestyle{plain} \setcounter{page}{1}


\section{Introduction and outline}
Since the original proposal by Bekenstein \cite{Bekenstein:1973ur}
that a black hole carries an entropy proportional to its area and
later discovery of the Hawking radiation \cite{Hawking:1974sw},
the microscopic origin of the thermodynamical behavior of black
holes has been one of the main puzzles of theoretical physics and
possibly one of the key problems for the understanding of quantum
gravity. The literature is really vast and here we will give an
account of just the references strictly relevant to our context.

A very simple but nonetheless instructive model to address the
problem of black hole entropy is the so called ``brick wall'' by
't Hooft \cite{'tHooft:1984re}. 't Hooft considers a quantum field
in the background of a classical black hole and using the WKB
approximation he derives the thermal entropy of the field outside
the horizon of the black hole. In performing the computation, two
spatial cutoffs are employed: a large distance one, needed to
avoid large volume divergences in the asymptotically flat region,
and a short distance one, the ``brick wall'', localized just
outside the horizon and suppressing the divergences due to the
growing number of modes close to the horizon. On the boundaries of
the space slice arising in this way, Dirichlet boundary conditions
are imposed. As noted in \cite{Mukohyama:1998rf,Mukohyama:1998ng},
the finite temperature state used in this quantization is a
thermal excitation of the Boulware vacuum \cite{Boulware:1974dm}.
The entropy obtained in this model is divergent in the limit of
vanishing brick wall thickness. These divergences were later
recognized as quantum corrections to the Bekenstein-Hawking
formula which can be absorbed into renormalization of the one loop
effective gravitational lagrangian
\cite{Susskind:1994sm,Demers:1995dq,Fursaev:1994ea,Cognola:1995km,deAlwis:1995cr,winstanley_01}
(see also \cite{Brustein:2005vx} for a recent perspective). Within
this context the ``brick wall'' takes the role of a useful
mathematical tool to regularize the theory. An interesting
different interpretation has been recently proposed in
\cite{Barbon:2003aq,Solodukhin:2005qy}.

The introduction of the brick wall cutoff as a more physical
device can be considered consistent with recent proposals arising
in string theory in which the nature of space-time ``inside'' the
horizon has deep quantum mechanical structure (see
\cite{Mathur:2005zp}).

In this paper we revisit the brick wall model. We first consider
in detail the case of 1+1 dimensional Schwarzschild black hole,
because it can be solved exactly without resorting to the WKB
approximation. We then argue that our results can be extended to
higher dimensions. In order to clarify the role of the boundary
conditions, we adopt a generic mixed (Robin) boundary condition,
which is the most general linear and local one. The cyclic state
used in the quantization is a Kubo-Martin-Schwinger (KMS)
state\footnote{See e.g. \cite{haag,brattellirobinsonII}}, with
respect to the Schwarzschild time. It accounts for the thermal
excitations over the Boulware vacuum. The Boulware vacuum
polarization can be exactly calculated in 1+1 dimensions, via the
energy-momentum conservation and the trace anomaly
\cite{Mukohyama:1998ng}. At this point one can derive the energy
inside the ``shell'' between the brick wall and any fixed point
outside the horizon. In the limit of vanishing brick wall
thickness, the energy of the shell is divergent unless we choose
the KMS state temperature $T$ to coincide with the Hawking
temperature $T_H$. In other words, the brick wall can be removed
only provided that $T=T_H$. This somewhat expected result does not
depend on the boundary conditions being in this sense universal
\cite{Hartle:1976tp,Jacobson:1994fp}.

In all the above considerations one has to keep in mind that the
expectation value of the energy-momentum tensor is a distribution
rather than a function. We observe in this respect that when
considering the horizon as part of the space-time in exam, the
above expectation value admits in general a non-vanishing Dirac
delta contribution, localized on the horizon. To our knowledge
this term has not been previously taken into account and indeed
appears in our computation in the limit of vanishing brick wall
thickness. We will show that it will also lead to a corresponding
surface term in the entropy density.

The entropy of a thermodynamical system is determined by its
energy up to an arbitrary constant. In most of the cases this
constant is fixed via the third principle of thermodynamics
(Nernst theorem), which requires that the entropy vanishes at zero
temperature. In our case, the computation of the entropy from the
energy of a shell outside the horizon is however a subtle matter.
The derivation of the entropy for generic $T$ must be performed
before removing the brick wall, since the removal is possible only
for $T=T_H$. Any attempt to determine at this stage the arbitrary
constant by the third principle, leads to divergences when later
removing the brick wall. These divergences are of the same kind
encountered in usual computation in the brick wall model. We can
get rid of them if we do not ask the quantum fields on such a
background to satisfy the third principle. We note here that the
issue on the third principle is a priori non related to the
observations regarding extremal black holes and Nernst theorem
(see for example \cite{Wald:1997qp,Belgiorno:2002pm}). Note also
that abandoning the third principle turns out to be a different
kind of regularization of the theory with respect to the one
previously considered in the literature
\cite{Demers:1995dq,Fursaev:1994ea,Cognola:1995km,deAlwis:1995cr,winstanley_01,Brustein:2005vx}
and involving infinite renormalization of the coupling constants.

Our calculations yield also a surface term for the entropy density
localized on the horizon. In our setting this term, together with
the corresponding surface term in the energy density, is
understood as a boundary effect: taking the brick wall boundary as
a pure mathematical tool, it can be considered as a (potentially
finite) contribution to the one loop renormalization of Newton
constant; on the other hand, taking the brick wall and the
boundary condition as somewhat more physical, we can interpret
these terms as an indication of the presence of physical degrees
of freedom localized on the horizon. We leave some more detailed
discussion on this for the last section of the paper.

The paper is organized as follows. The model is introduced in the
next section. In Sect. 3 we show that the total energy of a shell
outside the horizon is finite even in the limit of vanishing brick
wall thickness, provided that the temperature of the thermalized
field equals $T_H$. We also show the appearance of a surface term
in the energy density. In Sect. 4 we derive the entropy and
discuss the issue related to the third principle of
thermodynamics. Sect. 5 is devoted to a comparison with 't Hooft
results in the WKB approximation. In Sect. 6 we look for a
possible generalization to more realistic models in higher
dimensions. Finally, Sect. 7 contains a further discussion and our
conclusions.

\section{The model}\label{statement}

We consider a free real scalar field in a generic $n+1$
dimensional space-time $\{\mM,\, g\}$ with a time-like boundary
$\der\mM$. The action
\begin{equation}
\label{azioneSULbd} S=\int_\mM
\dd^{n+1}x\,\sqrt{|g|}\left(\frac{1}{2}\, g^{\mu\nu}\der_\mu
\varphi \der_\nu \varphi - \frac{1}{2} m^2 \varphi^2\right) -
\int_{\der \mM} \dd^n x\,\sqrt{|g_{\mathrm{ind}}|} \frac{\eta}{2}
\varphi^2 \, ,
\end{equation}
where $|g_{\mathrm{ind}}|$ is the determinant of the induced
(lorentzian) metric on $\der\mM$, implies both the Klein Gordon
equation
\begin{equation}
\label{KG} (g^{\mu\nu}\nabla_\mu\nabla_\nu + m^2) \vf=0
\end{equation}
and the Robin boundary condition
\begin{equation}
\label{condbordo} (g^{\mu\nu}N_\mu\der_\nu\vf-\eta\vf)|_{\der \mM}
= 0\, ,
\end{equation}
where $N^\mu$ is the unit vector normal to $\der\mM$. Using the
boundary condition \eqref{condbordo}, one can also express the
action as
\begin{equation}
\label{azioneCONbdINTERNA} S=\int_\mM
\dd^{n+1}x\,\sqrt{|g|}\left[\frac{1}{2}\,g^{\mu\nu} \der_\mu
\varphi \der_\nu \varphi - \frac{1}{2} m^2
\varphi^2-\frac{1}{2}g^{\mu\nu}\nabla_\mu(\vf\,\der_\nu
\vf)\right]\, .
\end{equation}
The stationarity condition for \eqref{azioneCONbdINTERNA} and the
boundary condition \eqref{condbordo} are equivalent to the
stationarity condition for the action \eqref{azioneSULbd}. We note
that the second expression for the action is suitable also for the
Dirichlet boundary condition. From \eqref{azioneCONbdINTERNA}
one derives the energy-momentum tensor
\begin{equation}
\label{tmnINbd}
\theta_{\mu\nu}(x)=\frac{2}{\sqrt{|g|}}\frac{\delta
S}{\delta\left[ g^{\mu\nu}(x)\right]} =
-\vf\nabla_\mu\nabla_\nu\vf+ \frac{1}{2}g_{\mu\nu}
\left(g^{\rho\sigma}\vf\nabla_\rho\nabla_\sigma\vf+m^2\vf^2\right)
\, .
\end{equation}

Let us focus now on a massless ($m=0$) scalar field in the
background of a $1+1$ dimensional classical black hole, with
metric
\begin{equation}
\label{1+1 metric} \dd s^2=f(r)\dd t^2 -\frac{1}{f(r)}\dd r^2 \, .
\end{equation}
We require asymptotic flatness
\begin{equation*}
\lim_{r\to \infty} f(r) = 1
\end{equation*}
and assume that $f$ has one and only one zero in $r=r_0$ (the
horizon) with positive \emph{surface gravity}
\begin{equation}
\kappa_0 \equiv \frac{1}{2}f'(r_0) > 0 \, . \label{surfgrav}
\end{equation}
We are thus considering a non-extremal black hole.

Following \cite{'tHooft:1984re}, we insert a brick wall at $r=\rho
> r_0$ and study the region $r\geq\rho$, which represents a static
space-time with time-like boundary.  We consider there the dynamics
with respect to the time-translation defined by the flux of
$\frac\der{\der t}$, the time-like Killing vector of the metric.
This is the Schwarzschild time for our model.

The classical equation of motion for the scalar field is
\begin{equation}
      \frac 1 {f(r)}\der^2_t\vf-\der_r\left [f(r)\der_r\vf\right ] =0\, .
\end{equation}
The Robin boundary condition \eqref{condbordo} takes the form
\begin{equation}
\label{condbordo1+1}
\left.\left(\sqrt{f(r)}\,\der_r\vf-\eta\vf\right)\right|_{r=\rho}=0
\, ,
\end{equation}
where $\eta $ in general can depend on $\rho$:
\begin{equation*}
\eta=\eta(\rho) \, .
\end{equation*}
In the coordinates $(t,y)$, with the ``tortoise coordinate" $y$
defined by
\begin{equation*}
y=y(r)\, , \qquad\frac{\dd y}{\dd r}=\frac{1}{f(r)} \, ,
\end{equation*}
the metric \eqref{1+1 metric} is conformally flat. Notice that
$f(r_0)=0$ and $f'(r_0)=2\kappa_0 > 0$ imply that for $r\tend
r_0$, to leading order
\begin{equation}
y(r) \approx \frac{1}{2\kappa_0}\ln 2\kappa_0(r-r_0)\, , \qquad
f(r) \approx \e^{2\kappa_0 y(r)}\, .
\end{equation}
In the coordinates $(t,y)$ the Klein-Gordon equation assumes its
flat space form
\begin{equation}
\label{eqm}
(\der_t^2-\der_y^2)\vf=0 \, ,
\end{equation}
and one is left with the problem of a scalar field on the
half-line $y>Y\equiv y(\rho)$ with the boundary condition
\begin{equation}
\label{condbordoY}
\left.\left(\der_y\vf-\nr\vf\right)\right|_{y=Y}=0\, , \qquad
\nr=\eta\sqrt{f(\rho)}\, .
\end{equation}
In order to avoid imaginary energies we ask
\begin{equation}
     \eta\geq0\, .
\end{equation}
The quantization of \eqref{eqm},\eqref{condbordoY} is easily
performed. The initial conditions are fixed by the canonical
equal-time commutation relations
\begin{equation}
\label{ccr} [\vf(t,y_1)\, ,\, \vf(t,y_2)]= 0 \, , \qquad
[\der_t\vf(t,y_1)\, ,\, \vf(t,y_2)]=-i\delta(y_1-y_2) \, .
\end{equation}
We introduce a class of quasi-free states $G_{\beta,\rho}$. Their
two point functions satisfy the Kubo-Martin-Schwinger condition.
They are Gibbs states at temperature $T=\beta^{-1}$ for
our model with brick wall position $\rho$. The relative
expectation values are denoted by $\langle\vf(t_1,y_1)\cdots
\vf(t_n,y_n)\rangle_{\beta,\rho}$. The basic one is the two-point
function \cite{Liguori:1996xr, Mintchev:2004jy}
\begin{eqnarray}
\label{twopoint}
\langle\vf(t_1,y_1)\vf(t_2,y_2)\rangle_{\beta,\rho} =
\qquad \qquad \qquad \qquad \qquad \nonumber \\
\int_\RR\frac{\dd p}{4\pi} \frac{|p|_\ell^{-1}}
{\e^{\beta |p|}-1}
\left[\e^{\beta |p|}\e^{-\ii |p|(t_1-t_2)}+\e^{\ii |p|(t_1-t_2)}\right]
\left[\e^{-\ii p (y_1-y_2)}+B(p,\rho)\e^{\ii
p(y_1+y_2-2Y)}\right]\, ,
\end{eqnarray}
where $B(p,\rho)$ is the reflection factor from the boundary
\begin{equation}
B(p,\rho)= \frac{p-\ii\nr}{p+\ii\nr} \label{rfactor}
\end{equation}
and the distribution $|p|_\ell^{-1}$ is defined by
\begin{equation}
|p|_\ell^{-1} \equiv \frac{\dd}{\dd p}\, \varepsilon(p) \ln (|p|
\ell )\, , \label{distribution}
\end{equation}
$\varepsilon$ being the sign function. The derivative in
\eqref{distribution} is understood in the sense of distributions.
The scale parameter $\ell$ has a well-known infrared origin
\cite{Grignani:1988fx}. Note that
\begin{equation}
\label{p ell property}
p\, |p|_\ell^{-1} = \varepsilon (p)\, ,
\end{equation}
which implies, as we shall see later on,
that $\ell$ is irrelevant in the calculation of the energy density.

The states $G_{\infty,\rho}$ can be considered as the analogous of
the Boulware vacuum. Indeed they appear as usual vacuum to an
observer at rest in the $r$ coordinate in the asymptotically flat
region. They are annihilated by every destruction operator
associated to a normal mode with respect to the Schwarzschild
time.

\section{Energy}\label{energy}

In this Section we discuss in detail the derivation of the energy
density for the states introduced in Section 2. At the Hawking
temperature, the divergences occurring in the thermal energy of a
shell outside the horizon are perfectly balanced by the Boulware
vacuum polarization. From general considerations on the structure
of the stress energy tensor, we cannot exclude a contribution to
the energy density in the form of a surface term localized on the
horizon. The introduction and following removal of a brick wall as
a regularization confirms the presence of a finite term of this
form. In general, the Boulware vacuum polarization could give a
further surface contribution which cannot be determined in this
setting. In principle a complete determination of such a term can
be performed by some sort of measurement or deduced by a better
understanding of this semiclassical picture in the context of a
quantum theory of both matter and gravity.

\subsection{Wald's axioms and definition}
Given one of the Gibbs states described in the last section, we
want to calculate the expectation value of the thermal excitations
of the energy-momentum tensor over the Boulware-like vacuum. Wald
showed \cite{Wald:1977up,Wald:1978pj,wald} that a correctly
renormalized energy-momentum tensor $T^\mu_{\phantom\mu\nu}$
obeying certain assumptions is essentially unique. Wald's
requirements are
\begin{enumerate}
     \item \emph{Conservation}. Given any state $\alpha$
     \begin{equation}
       \nabla_\mu \langle T^\mu_{\phantom\mu\nu}\rangle_\alpha = 0\, .\label{e-m
tensor conservation}
     \end{equation}
     \item \emph{Consistency}. Given any regular enough couple of
     states $\alpha_1,\alpha_2$ we have that
     $\langle T^\mu_{\phantom\mu\nu}\rangle_{\alpha_1}-\langle
T^\mu_{\phantom\mu\nu}\rangle_{\alpha_1}$
     is defined by the usual point-splitting procedure.
     \item \emph{Causality holds} in the form of a locality
     requirement.
     \item \emph{Normalization}. In Minkowski space-time, being
     $\Omega$ the usual Fock vacuum,
     $\langle T^\mu_{\phantom\mu\nu}\rangle_\Omega=0$.
\end{enumerate}
In a generic $n+1$ dimensional space-time, from the contraction of
a Killing vector $K$ with $T^\mu_{\phantom{\mu}\nu}$ we can
construct a form $J_K$ whose expectation values satisfy the
following relations
\begin{gather}
     \langle J_K\rangle=\sqrt{|g|}\epsilon_{\mu_1\cdots\mu_n\nu}\langle
T^\nu_{\phantom\nu \rho}\rangle K^\rho\dd
     x^{\mu_1}\wedge\cdots\wedge\dd x^{\mu_n}\, ,\qquad\\
     \dd \langle J_K\rangle=0\, .
\end{gather}
In our model we consider the Killing vector $K=\der_t$. For any
state $\alpha$, integrating the second relation above, we can
define, the \emph {energy} inside a ``shell'' (actually a segment)
$(r_1,r_2)$
\begin{equation}\label{first energy in a shell}
    E(r_1,r_2) = \int_{r_1}^{r_2} \langle T^t_{\phantom t
t}(t,r)\rangle_\alpha\dd r\, .
\end{equation}
We define
\begin{gather}\label{point splitting definition}
\langle\theta^\mu_{\phantom{\mu}\nu}(x)\rangle_{\beta,\rho}\equiv
\qquad \qquad \qquad \qquad \qquad \qquad \\\notag \lim_{x'\to x}
\left(-\nabla^\mu\nabla_\nu+ \frac{1}{2}g^\mu_{\phantom{\mu}\nu}
g^{\rho\sigma}\nabla_\rho\nabla_\sigma \right) \left
[\langle\vf(x')\vf(x)\rangle_{\beta,\rho}-
\langle\vf(x')\vf(x)\rangle_{\infty,\rho}\right ]\, .
\end{gather}
Then, the second of Wald's requirements implies
\begin{equation}\label{second Walds for us}
     \langle\theta^\mu_{\phantom{\mu}\nu}\rangle_{\beta,\rho}=\langle
T^\mu_{\phantom{\mu}\nu} \rangle_{\beta,\rho}-
     \langle T^\mu_{\phantom{\mu}\nu}
\rangle_{\infty,\rho}
\end{equation}
and thus the energy inside a shell for the Gibbs state
at temperature $\beta$ is given by
\begin{equation}\label{energy inside a shell definition}
E_{\beta,\rho}(r_1,r_2)=\int_{r_1}^{r_2}\left[\langle\theta^t_{\phantom
t t}(r)\rangle_{\beta,\rho}+\langle T^t_{\phantom t
t}(r)\rangle_{\infty,\rho}\right]
      \dd r\, .
\end{equation}
In view of the point-splitting procedure \eqref{point splitting
definition}, the expression \eqref{twopoint} for the two point
function and the property \eqref{p ell property}, an integration
by parts and a change of variable give
\begin{equation}
     E_{\beta,\rho}(r_1,r_2)=\int_{y(r_1)}^{y(r_2)}\eps_{\beta,\rho}(y)
\dd y+\int_{r_1}^{r_2}\langle T^t_{\phantom t
t}(r)\rangle_{\infty,\rho}\dd r\,,
\end{equation}
where
\begin{equation}\label{epsilon beta rho}
\eps_{\beta,\rho}(y)= \int_\RR\frac{\dd p}{2\pi}\frac{|p|}
{\e^{\beta|p|}-1}\left [1+ B(p,\rho) \e^{2\ii p(y-Y)}\right ]\, ,
\end{equation}
which can also be expressed in a manifestly real form as
\begin{equation*}
\eps_{\beta,\rho}(y) = \int_0^\infty\!\frac{\dd
p}{\pi}\frac{p}{\e^{\beta p}-1} \left\{
1+\frac{p^2-\nr^2}{p^2+\nr^2}\cos [2p(y-Y)]+
\frac{2p\nr}{p^2+\nr^2}\sin [2p(y-Y)]\right \}\, .
\end{equation*}
We can single out the usual Stefan-Boltzmann contribution to the
thermal energy and a specific contribution due to the boundary
\begin{gather}
     \eps_{\beta,\rho}(y) = \frac\pi{6\beta^2} +
     h_{\beta,\rho}(y-Y)\, ,\\
h_{\beta,\rho}(\xi)\equiv
\frac{1}{\beta^2}\mathcal{F}\left[\frac{|p|}{(\e^{|p|}-1)}
\frac{(p-\ii\nr\beta )}{(p+\ii\nr\beta )}\right]
\left(\frac{2\xi}{\beta}\right) \, ,
\end{gather}
$\mathcal{F}$ being the Fourier transform
\begin{equation*}
\mathcal{F}[g(p)](x)=\int_{-\infty}^{+\infty} \frac{\dd
p}{2\pi}g(p)\,\e^{\ii p x}\, .
\end{equation*}
We note that since the function
$\frac{|p|}{(\e^{|p|}-1)}\frac{(p+\ii\nr\beta )}{(p-\ii\nr\beta
)}$ is continuous and $L^1$, its Fourier transform is continuous,
$L^1$ and infinitesimal at infinity.

\subsection{Vacuum polarization and boundary contribution}

In order to get a complete expression for the energy in
\eqref{energy inside a shell definition} we still have to
determine $\langle T^\mu_{\phantom\mu\nu}\rangle_{\infty,\rho}$,
that is the Boulware vacuum polarization. In our 1+1 dimensional
model the expectation value $\langle
T^\mu_{\phantom\mu\nu}\rangle$ for any state that does not imply
transport, i.~e. $\langle T^t_{\phantom t r}\rangle=0$, is almost
completely determined \cite{Mukohyama:1998ng} by its conservation
law \eqref{e-m tensor conservation} and the trace anomaly
\begin{equation}
     \langle T^\mu_{\phantom\mu\mu}\rangle_{\beta,\rho}=\frac1{24\pi}R \, ,
\label{e-m tensor trace anomaly}
\end{equation}
where $R$ is the scalar curvature and in our case $R=f''$.
The integration gives
\begin{equation}\label{1+1 integra trr}
      f(r)\langle T_{\phantom r r}^r(r)\rangle=\frac1{24\pi}\kappa^2(r)-C\, ,
\end{equation}
where $C$ is an integration constant and
$\kappa(r)=\frac{1}{2}f'(r)$. Since this is an equation involving
distributions, the general solution is given by
\begin{equation*}
     \langle T_{\phantom r
r}^r(r)\rangle=\frac1{f(r)}\left[\frac1{24\pi}\kappa^2(r
)-C\right]-U\,\delta(r-r_0)\, ,
\end{equation*}
where $U$ is an arbitrary constant with the dimensions of an
energy. We can call the $U \delta$ term a ``boundary'' term.
Different values of the constants $C$ and $U$ identify
different states. By means of the trace anomaly one thus derives
\begin{equation}\label{1+1 ttt rin con costanti arbitrarie}
\langle T^t_{\phantom t
t}(r)\rangle=\frac1{f(r)}\left[C-\frac1{24\pi}\kappa^2(
r)\right]+\frac1{24\pi}f''(r)+U\delta(r-r_0)\, .
\end{equation}
As we noted in the previous section, for any $\rho$, the Boulware
like state appears as vacuum to an observer in the asymptotically
flat region; since $\kappa(r)\tend 0$ when $r\tend\infty$, it is
identified by the choice $C=0$. Moreover, since we are dealing
with the $r\geq\rho>r_0$ region, the ``boundary'' term is
irrelevant. We can thus write
\begin{equation*}
\langle T^t_{\phantom t t}(r)\rangle_{\infty,\rho}=
-\frac{1}{24\pi}\frac{\kappa^2(r)}{f(r)}+\frac{1}{24\pi}f''(r)\, .
\end{equation*}
Note that it is actually $\rho$ independent.

We now write the energy in a segment or ``shell'' $[r_1,r_2]$ as
\begin{gather}
E_{\beta,\rho}(r_1,r_2)= \int_{r_1}^{r_2}\langle T^t_{\phantom t
t}(r)\rangle_{\beta,\rho} \dd
r=E^{(b)}_{\beta,\rho}(r_1,r_2)+E^{HH}_{\beta,\rho}(r_1,r_2)\, ,\\
E^{(b)}_{\beta,\rho}(r_1,r_2)
=\int_{y(r_1)}^{y(r_2)}h_{\beta,\rho}(y-Y)\, ,\\
E^{HH}_{\beta,r_0}(r_1,r_2)=\int_{r_1}^{r_2}\!\left [\frac\pi{6\beta^2}\frac
1{f(r)}+ \langle T^t_{\phantom t t}(r)\rangle_{\infty,\rho}\right ]\, .
\end{gather}
We will show in the following that the first term can be regarded
as a pure boundary contribution while the second term, with the
appropriate choice for $\beta$, can be regarded as the
contribution from the ``Hartle-Hawking'' state
\cite{Hartle:1976tp,Jacobson:1994fp}.

\subsection{Removing the brick wall}

Now we study the limit $\rho\tend r_0$, that is the limit of
vanishing brick wall thickness.
We first consider the case of fixed $r_1$ and $r_2$ satisfying
$r_2>r_1>\rho$. As $\rho\tend r_0$,
$Y\tend - \infty$ and so, as noted at the end of Section 3.1,
$h_{\beta,\rho}(y-Y)\tend 0$ pointwise inside all of the segment,
and we get
\begin{equation}\label{Er1r2}
\begin{split}
&E^{(b)}_{\beta,r_0}(r_1,r_2)=0\, ,\\
&E^{HH}_{\beta,r_0}(r_1,r_2)=\int_{r_1}^{r_2}\!\left [\frac\pi{6\beta^2}\frac
1{f(r)}+ \langle T^t_{\phantom t t}(r)\rangle_{\infty,r_0}\right ]
\dd r.
\end{split}
\end{equation}

In the case of the segment $r_1=\rho,r_2=\sigma$ with $\sigma$
independent on $\rho$, we have in the limit \mbox{$\rho\tend r_0$}
\begin{multline}\label{energy boundary term definition}
E^{(b)}_{\beta,\rho}(\rho,\sigma)=\int_Y^{y(\sigma)}h_{\beta,\rho}(y-Y)\dd
y=\\=\int_0^{y(\sigma)-Y}\!
     h_{\beta,\rho}(\xi)\dd
     \xi  \tend\int_0^\infty\!
     h_{\beta,r_0}(\xi)\dd \xi=E^{(b)}_\beta\, ,
\end{multline}
where $E^{(b)}_\beta$ depends only on $\beta$ and
$\eta_{r_0}=\lim_{\rho\tend r_0}\eta_\rho$:
\begin{multline}
E^{(b)}_{\beta}=\frac{1}{\beta^2}\int_0^\infty\!\mathcal{F}\left[\frac
{\abs{p}}{\e^{|p|}-1}\frac{p-\ii\eta_{r_0}\beta}{p+\ii\eta_{r_0}\beta}
\right]\left(\frac{2\xi}{\beta}\right)\dd
\xi=
\\
=\frac1\beta\int_0^\infty\!\mathcal{F}\left[\frac{|p|}{\e^{|p|}-1}
\frac{p-\ii\eta_{r_0}\beta}{p+\ii\eta_{r_0}\beta}\right]\!(2\xi)\dd \xi.
\end{multline}
    From equations \eqref{Er1r2},\eqref{energy boundary term
definition} we can deduce that the function $h_{\beta,\rho}$
determines a pure boundary term in the stress energy tensor localized
on the horizon.

We consider now
\begin{multline}
      E^{HH}_{\beta,\rho}
      (\rho,\sigma)=\int_\rho^\sigma\!\left
[\frac\pi{6\beta^2}\frac1{f(r)}+\langle
T^t_{\phantom t t}(r)\rangle_{\infty, \rho}\right ]\dd r=
\\
\int_\rho^\sigma\!\left [\frac\pi{6\beta^2}\frac1{f(r)}-\frac{1}{24\pi}
\frac{\kappa^2(r)}{f(r)}+\frac{1}{2 4\pi}f''(r)\right ]\dd r\, .
\end{multline}
This term may diverge when $\sigma\tend\infty$ or $\rho\tend r_0$.
The first one is a well understood volume divergence and it is not
interesting to us. We thus keep $\sigma$ constant and finite.
Then,
\begin{multline}\label{energia HH sviluppata}
      E^{HH}_{\beta,\rho}
(\rho,\sigma)=\left(\frac\pi{6\beta^2}-\frac{\kappa_0^2}{24\pi}\right)
X(\rho,\sigma)+
\frac{\kappa_0^2}{24\pi}\Delta(\rho,\sigma)+\frac{1}{12\pi}
\left [\kappa(\sigma)-\kappa(\rho)\right ]\, ,
\end{multline}
where
\begin{gather}
     X(r_1,r_2)\equiv \int_{r_1}^{r_2}\!\frac 1 {f(r)} \dd
     r=y(r_2)-y(r_1)\, , \\
\int_{r_1}^{r_2}\!\frac 1 {24\pi} \frac {\kappa(r)^2} {f(r)} \dd r
= \frac {\kappa_0^2}{24\pi} \left [X(r_1,r_2)-\Delta (r_1,r_2)\right ]
\end{gather}
    In the limit $\rho\approx r_0$,
\begin{gather}
     X(\rho,\sigma)\approx\frac 1 {2\kappa_0}\ln\frac a {\rho-r_0}\, ,\\
     \Delta (\sigma,\rho)\quad \text{is finite}\, ,
\end{gather}
where $a$ depends on $\sigma$. The quantity in \eqref{energia HH sviluppata}
is thus divergent in
the limit $\rho\tend r_0$, unless we put
\begin{equation}
\beta=\beta_H=\frac1{T_H}=\frac{2\pi}{\kappa_0}.
\end{equation}
In this case the divergences due to thermal excitations are
perfectly balanced by the Boulware vacuum polarization and we get
a finite energy for every $\sigma$. This can be considered as a
definition of the Hawking temperature $T_H$ for our model.

At $T=T_H$ it is thus possible to remove the brick wall and we are
left with the expression:
\begin{multline}
      \langle T_t^t\rangle_{\beta_H,
r_0}=\frac1{f(r)}\left[\frac\pi{6\beta_H^2}-\frac{\kappa^2(r)}{24\pi}\right]+
\frac{f''(r)}{24\pi}+E^{(b)}_{\beta_H}\delta(r-r_0)=\\
      =\langle T^t_t\rangle_{HH}+E^{(b)}_{\beta_H}\delta(r-r_0)\, ,
\end{multline}
where $\langle T^\mu_{\phantom\mu \nu}\rangle_{HH}$ is known as
the expectation value of $ T^\mu_{\phantom\mu \nu}$ in the Hartle
Hawking
state\cite{Hartle:1976tp,Jacobson:1994fp,sew80,Fredenhagen:1989kr,wald}.
Indeed, by definition, the expectation value of the
energy-momentum tensor in the Hartle Hawking state is regular on
both the past and the future horizon of the black hole; this
corresponds to the choice
\begin{equation*}
C=\frac\pi6\left(\frac{\kappa_0}{2\pi}\right)^2 =
\frac\pi{6\beta_H^2}
\end{equation*}
in  equation \eqref{1+1 ttt rin con costanti arbitrarie}.
To an observer at rest in the asymptotically flat region, the
Hartle Hawking state appears as a thermal bath at the Hawking
temperature $T_H$.

\section{Entropy}\label{entropy}

In this Section we describe the derivation of the entropy of a
shell of our space-time from the previously calculated energy
density. Even at the Hawking temperature, the entropy in a shell
attached to the horizon is divergent. The divergence can be
resolved allowing that the system of the quantum field in this
background does not satisfy Nernst theorem. The price for this is
the presence of an arbitrary constant in the entropy of any shell.
One can consider this freedom as a consequence of a breakdown of
the semiclassical picture or as an intrinsic feature of the QFT in
this background (analogous to renormalization effects), or as a
combination of both.

The entropy density admits a surface term localized on the
horizon. It indicates the presence of physical degrees of freedom
there, which can give rise to a (potentially finite)
renormalization of Newton constant.

\subsection{Entropy from energy}

We will now perform the calculation of the entropy inside a shell. Since
\begin{equation*}
      E=\frac{\der(\beta F)}{\der\beta}\, ,\qquad F=E-S/\beta\, ,
\end{equation*}
     we have
\begin{equation}\label{energia entropia generiche F E}
      S(\beta)=\beta
E(\beta)-\bar{\beta}F(\bar{\beta})-\int_{\bar{\beta}}^\beta\!E(b)\dd b\, .
\end{equation}
When the relations
\begin{equation}\label{1+1 ipotesi terzo principio}
      E(\beta)\tend E_0\,, \quad \beta [E(\beta)-E_0]\tend 0
\qquad\text{for}\quad \beta \tend \infty
\end{equation}
are satisfied, the arbitrary constant $\bar{\beta}F(\bar{\beta})$
can be determined via the third principle of thermodynamics in the form
\begin{equation*}
S(\beta)\tend0\qquad\text{for}\quad \beta\tend \infty \, ,
\end{equation*}
that gives
\begin{equation*}
F(\bar{\beta})=-\frac{1}{\bar{\beta}}\int_{\bar{\beta}}^\infty\!(E(b)-
E_0)\dd b+E_0\, .
\end{equation*}
It is straightforward to show that the conditions in \eqref{1+1
ipotesi terzo principio} are necessary and sufficient for the
system to verify the third principle.

\subsection{Removing the brick wall with and without third
principle}

Since the removal of the brick wall is not possible
``off shell'', i.e. for $\beta\neq\beta_H$, we are forced to
perform the integration in \eqref{energia entropia generiche F E}
at non zero brick wall thickness and then to perform the limit
$\rho\tend r_0$.

Let's consider the energy in a shell $(r_1,r_2)$. The analysis of
the previous section gives
\begin{multline*}
E_{\beta,\rho}(r_1,r_2)=\int_0^{X(r_1,r_2)}\!h_{\beta,\rho}(\xi)\dd\xi+
\left(\frac\pi{6\beta^2}-\frac{\kappa_0^2}{24\pi}\right)X(r_1,r_2)+\\+
\frac{\kappa_0
^2}{24\pi}\Delta(r_1,r_2)+
      \frac{1}{24\pi}\left[2\kappa(r_2)-2\kappa(r_1)\right]\, .
\end{multline*}
We recall that
\begin{equation*}
h_{\beta,\rho}(\xi) =
\frac{1}{\beta^2}\mathcal{F}\left[\frac{|p|}{(\e^{|p|}-1)}
\frac{(p-\ii\nr\beta )}{(p+\ii\nr\beta )}\right]
\left(\frac{2\xi}{\beta}\right)\, .
\end{equation*}
Using equation \eqref{energia entropia generiche F E} and putting
\mbox{$\bar{\beta}=\beta_H=\frac{2\pi}{\kappa_0}$} we have
\begin{multline*}
S_{\beta,\rho}(r_1,r_2)=\frac{\pi}{6\beta}X(r_1,r_2)+
\beta\int_0^{X(r_1,r_2)}\!h_{\beta,\rho}(\xi)\dd\xi+\\+ \beta
_H\frac{1}{24\pi}\left[-\kappa_0^2
X(r_1,r_2)+\kappa_0^2\Delta(r_1,r_2)+2\kappa(r_2)-2\kappa(r_1)\right]+\\
-\beta_H F_H-X(r_1,r_2)\int_{\beta_H}^\beta\!\frac{\pi}{6b^2}\dd
b-\int_{\beta_H}^\beta \!\int_0^{X(r_1,r_2)}\!h_{b,\rho}(\xi)\dd b
\dd \xi.
\end{multline*}
The third principle is satisfied if
\begin{multline*}
      \beta_H
      F_H=-X(r_1,r_2)\int_{\beta_H}^\infty\!\frac{\pi}{6b^2}\dd b
      -\int_{\beta_H}^\infty\!\int_0^{X(r_1,r_2)}\!h_{b,\rho}(\xi)\dd \xi
\dd b \,+\\
+\beta_H\frac{1}{24\pi}\left[-\kappa_0^2
X(r_1,r_2)+\kappa_0^2\Delta(r_1,r_2)+2\kappa(r_2)-2\kappa(r_1)\right]\, .
\end{multline*}
Hence, going ``on shell'', we have:
\begin{multline}\label{Sbhrs}
S_{\beta_H,\rho}(r_1,r_2)=\\
=\frac{\pi}{3\beta_H}X(r_1,r_2)+
\beta_H\int_0^{X(r_1,r_2)}\!h_{\beta_H,\rho}(\xi)\dd\xi+
\int_{\beta_H}^\infty\!\int_0^{X(r_1,r_2)}\!h_{b,\rho}(\xi)\dd\xi\dd b\, .
\end{multline}
We note here that the function
\begin{equation*}
\int_{\beta_H}^\infty\!h_{b,\rho}(\xi)\dd
b=-\frac{1}{\beta_H}\mathcal{F}\left[\ln
(1-\e^{-|p|})\frac{p-\ii\nr\beta_H}{p+\ii\nr\beta_H}\right]\left(\frac
{2\xi}{\beta_H}\right)
\end{equation*}
is not $L^1$ but it is infinitesimal at infinity, since it is the
Fourier transform of a non continuous $L^1$ function. One can
analyze \eqref{Sbhrs} in analogy of what we did in the previous
section for the energy.

In the limit $\rho\tend r_0$ with fixed $r_1,r_2$ satisfying
$r_2>r_1>\rho$, the second and third term in expression
\eqref{Sbhrs} vanish and we are left with an usual volume term:
\begin{equation}
     S_{\beta_H,\rho}(r_1,r_2)=\frac{\pi}{3\beta_H}X(r_1,r_2)\, .
\end{equation}

When $r_1=\rho$ and $r_2=\sigma$, with $\sigma$ constant and
finite, we have in the limit $\rho\tend r_0$
\begin{equation*}
      X(\rho,\sigma) \tend\infty\quad,\quad \Delta(\rho,\sigma) \,\text{is
      finite}\, .
\end{equation*}
The first term in \eqref{Sbhrs} is linearly divergent in $X$, and
the last one can also be divergent. The function
$\int_{\beta_H}^\infty\!h_{b,\rho}(\xi)\dd b$ is not $L^1$ but is
infinitesimal at infinity: the possible divergence from the last
term in \eqref{Sbhrs} can not be linear in $X$\footnote{In
principle, we can not be sure of the actual appearance of such a
divergence since the integral could be
convergent in improper way.}.\\
Its origin can be traced back to the $\beta$ dependence of the so
called ``surface'' term in the energy
\begin{equation*}
     E^{(b)}_{\beta,\rho}=\int_0^X\!h_{\beta,\rho}(\xi)\dd\xi=\int_0^X\!
     \frac1{\beta^2}\mathcal{F}\left[\frac{|p|}{\e^{|p|}-1}\frac{p-\ii\eta
_\rho\beta}{p+\ii\eta_\rho\beta}\right]\left(\frac{2\xi}{\beta}\right)
\dd\xi\, .
\end{equation*}
As we showed in the preceding section, in the  limit $\rho\tend
r_0$ we have
\begin{equation*}
E^{(b)}_\beta=\frac1\beta\int_0^\infty\!\mathcal{F}\left[\frac{|p|}{\e
^{|p|}-1}\frac{p-\ii\eta_{r_0}\beta}{p+\ii\eta_{r_0}\beta}\right]\!(2\xi)\dd
      \xi\, .
\end{equation*}
When $\beta\tend\infty$, or more precisely
$\beta\gg\eta_{r_0}^{-1}$, we have, up to leading order:
\begin{equation*}
E^{(b)}_\beta=-\frac1\beta\int_0^\infty\!\mathcal{F}\left[\frac{|p|}
{\e^{|p|}-1}\right]\!(2\xi)\dd
      \xi\propto\frac1\beta
\end{equation*}
and the $\beta$ dependence is incompatible with the third
principle. Such a behavior can be qualitatively understood looking
at the description of the system in the $(t,y)$ coordinates: when
$\rho\tend r_0$ we have $Y\tend -\infty$ and thus the specific
boundary condition cannot appear independently in the leading
order expansion of thermodynamical quantities. Due to the
conformal symmetry of our two dimensional model, the only
dimensional parameter relevant for the expansion is $\beta$ and
thus $\beta^{-1}$ is the only possible leading term.

The different behavior of the two divergences reflects their
different physical origin. The one linear in $X$ is basically due
to the thermal excitations of the model close to the horizon and
we expect it to be present also in more realistic models; the
other one can be regarded as a specific feature of our
oversimplified model.

We have thus shown that the entropy of a shell contains two
distinct terms:
\begin{gather}
S_{\beta_H,\rho}(r_1,r_2)=S_{\beta_H,\rho}^{HH}(r_1,r_2)+S_{\beta_H,\rho}^{(b)}(r_1,r_2)\\
S_{\beta_H,\rho}^{HH}(r_1,r_2)\equiv\frac{\pi}{3\beta_H}X(r_1,r_2)\\
S_{\beta_H,\rho}^{(b)}(r_1,r_2)\equiv\beta_H\int_0^{X(r_1,r_2)}\!h_{\beta_H,\rho}(\xi)\dd\xi+
\int_{\beta_H}^\infty\!\int_0^{X(r_1,r_2)}\!h_{b,\rho}(\xi)\dd\xi\dd
b\, .
\end{gather}
The first one is a volume term, which is also obtained in the
usual treatment of the brick wall model (see Section 5). The
second one is a surface term localized on the horizon which is
inherited from the surface term in the energy density.

Both these terms are divergent in the limit $\rho\tend r_0$ but
both divergences can be avoided if we do not use the third
principle in order to fix the arbitrariness in the determination
of the entropy. With the substitution
\begin{equation*}
      \beta_H F_H\tend\beta_H F_H+\frac{\pi}{3\beta_H}\frac{1}{2\kappa_0}X
      +\int_{\beta_H}^\infty\!\int_0^X\!h_{b,\rho}(\xi)\dd\xi\dd
b-S^\Omega_{\beta_H}(\rho,\sigma)
\end{equation*}
we have
\begin{equation}\label{SbetaHfinale}
S_{\beta_H,r_0}(r_0,\sigma)=\beta_H\int_0^\infty\!
\frac{1}{\beta_H^2}\mathcal{F}\left[\frac{|p|}{(\e^{|p|}-1)}
\frac{(p-\ii\eta_{r_0}\beta_H )}{(p+\ii\eta_{r_0}\beta_H )}\right]
\left(\frac{2\xi}{\beta_H}\right)\dd\xi+S^\Omega_{\beta_H}(r_0,\sigma)\, ,
\end{equation}
where $S^\Omega_{\beta_H}(\rho,\sigma)$ is an arbitrary continuous
function, which parameterize the inability of determining the
exact value of the entropy without any external information input
such as, for example, a derivation from a more fundamental theory
which correctly identifies the degrees of freedom of the system.
This picture recalls the standard situation one has in
renormalization theory, where we pay the finiteness of the theory
with an indetermination.

As we already noted, we expect that the divergences appearing in
the surface term of the entropy are a specific feature of our
extremely simplified model. Thus we expect that a surface term is
generically present and potentially finite, showing an
accumulation of degrees of freedom on the horizon which is
different from the one usually considered as a consequence of the
growing number of modes close to the horizon. This can be
interpreted as a first order quantum correction to the
Bekenstein-Hawking formula due to the interaction of the matter
field with the gravitational field. Also this term can be
considered as affected by some sort of indetermination since we
cannot exclude a priori the presence of a surface contribution
also in the arbitrary function $S^\Omega_{\beta_H}(r_0,\sigma)$.
In general, the boundary contribution is non-vanishing whenever
\begin{equation}
S_{\beta_H,r_0}(r_0,r_0)=\beta_H\int_0^\infty\!h_{\beta_H,\rho}(\xi)\d
d\xi+S^\Omega_{\beta_H}(r_0,r_0)\neq0\, .
\end{equation}

The expression for the entropy in a shell far outside the
horizon can be deduced from~\eqref{SbetaHfinale}
\begin{equation}\label{Sr1r2Omega}
S_{\beta_H,r_0}(r_1,r_2)=S^\Omega_{\beta_H}(r_0,r_2)-S^\Omega_{\beta_H
}(r_0,r_1)\, .
\end{equation}
If we assume that the origin of the awkward behavior of the
entropy is in some sort of interaction with the black hole and its
horizon, it seems reasonable to ask for the expression
\eqref{Sr1r2Omega} to become equal to the expected one in the
asymptotically flat region, that is
\begin{equation*}
S^\Omega_{\beta_H}(r_0,r_2)-S^\Omega_{\beta_H}(r_0,r_1)\tend\frac\pi{3
\beta}(r_2-r_1)\qquad
      \text{for}\quad r_1,r_2\tend\infty \, .
\end{equation*}

\section{The WKB approximation}\label{wkb}

It is instructive to reexamine the model described in the previuos
sections within the WKB approximation. For this purpose we
consider the shell $(\rho,\sigma)$ and define the wave
number $k(r)$
\begin{equation*}
      k(r)=\frac E{f(r)}\, .
\end{equation*}
The density of states in the WKB approximation is given by
\begin{gather*}
      n(E)=\frac{\der N(E)}{\der E}\, ,\\
        \pi N(E)=\int_\rho^\sigma\!\dd r \frac E{f(r)}\, .
\end{gather*}
We note that for Neumann or Dirichlet boundary conditions
in 1+1 dimensions the WKB approximation gives the right
eigenvalues of the energy. The free energy $F$ reads
\begin{equation*}
      F_{\rm WKB}=\int\!n(E) \ln(1-\e^{-\beta E})\dd
      E=\int\!\frac{N(E)}{\e^{\beta
      E}-1}=\frac\pi{6\beta^2}\int_\rho^\sigma\frac{\dd r}{f(r)}\, .
\end{equation*}
Recalling that $S=\beta^2\der_\beta F$ and neglecting subleading
terms in the limit of $\rho\approx r_0$, one obtains
\begin{equation*}
      S_{\rm WKB}=\frac{\pi}{3\beta_H}X(\rho,\sigma)=
S_{\beta_H,\rho}^{HH}(\rho,\sigma)\, .
\end{equation*}
Again neglecting subleading terms and comparing with \eqref{Sbhrs}
one finds
\begin{equation*}
       S_{\beta_H,\rho}(\rho,\sigma)=S_{\rm
WKB}+S^{(b)}_{\beta_H,\rho}(\rho,\sigma)\, ,
\end{equation*}
showing that the the boundary term in the entropy is lost in the
WKB approximation. In both cases however the
entropy is divergent in the limit $\rho\tend r_0$ and the
considerations at the end of the previous section apply.

\section{Higher dimensions}\label{highdim}

In the higher dimensional and massive extension of our model
we have to deal with two main issues:
\begin{itemize}
      \item resolving the spectral problem for the field, i.e. the
      determination of the its normal modes;
      \item determining the Boulware vacuum polarization.
\end{itemize}
These problems are not conceptual but technical. The following
steps towards their solution can be made in the spherically symmetric case.

The metric for an $n+1$ dimensional spherical black hole is given by
\begin{equation*}
      \dd s^2=U(r)\dd t^2-\frac1{U(r)}\dd
r^2+r^2\Omega^{(n-1)}_{ij}\dd\theta^i \dd\theta^j \, ,
\end{equation*}
where $U$ is assumed to have one and only one zero in $r_0$ and
\mbox{$U'(r_0)=2\kappa_0\neq0$}. We insert a brick wall and
consider a scalar field only in the exterior of the sphere of
radius $\rho=r_0$, with a boundary condition
\begin{equation*}
\left.  \left(\sqrt{U}\der_r\vf=\frac h r_0
\vf\right)\right|_{r=\rho}.
\end{equation*}
Using the ``tortoise'' coordinate $r^*$ defined by
\begin{equation*}
      \frac{\dd r^*}{\der r}=\frac1U
\end{equation*}
the thermal energy in a spherical shell is given by
\begin{equation*}
E_{\beta,\rho}=\int_{r_1}^{r_2}\!\langle\theta^t_{\phantom t
t}(r)\rangle_{\beta,\rho}\Sigma_{n-1}
      r^{n-1}\dd r=\int_{r^*_1}^{r^*_2}\! \eps(r^*)_{\beta,\rho}\Sigma_{n-1}
      (r(r^*))^{n-1}\dd r^*\, ,
\end{equation*}
where $\Sigma_{n-1}r^{n-1}$ is the area of the $(n-1)$ dimensional
sphere of radius $r$,
$\Sigma_{n-1}=\frac{2\pi^{n/2}}{\Gamma\left(\frac n2\right)}$.
At this point one can separate the angular dependence and pass to the reduced
radial wave functions $f=r^{\frac{n-1}2}\psi$, where $\psi$ is the
original radial wave function. Then one is left with the study of the
operators
\begin{equation*}
      D_l=-\der_{r^*}^2+\frac{n-1}{2r}U\left(\der_rU\right)+
      \frac{(n-1)(n-3)}{4}\frac{U^2}{r^2}+U\frac{l(l+n-2)}{r^2}+
      U m^2 \, ,
\end{equation*}
where $r=r(r^*)$, $l(l+n-2)$ is the eigenvalue of the angular part
of the flat Laplace operator in $n$ dimension (the squared total
angular momentum) and $m$ is the mass of the field. Now, one has to
investigate the spectral problem for $D_l$ with the boundary condition
\begin{equation*}
\left.\left(\frac{1}{\sqrt{U}}\der_{r^*}f=\frac{h+\frac{n-1}{2}}{r_0}f
\right)\right|_{r^*=\rho^*}\, ,
\end{equation*}
where $\rho^*=r^*(\rho)$. Being $f_{\lambda l}(r^*)$ the complete set of
orthonormal (improper) eigenfunctions, i.~e.
\begin{gather*}
D_l f_{\lambda l}=\lambda^2 f_{\lambda _l}\, ,\\
      \int\!\sigma_{\lambda l}f_{\lambda l}(r_1^*)f_{\lambda
      l}(r_2^*)\dd \lambda=\delta(r_1^*-r_2^*)\, ,
\end{gather*}
we get
\begin{equation}
\eps(r^*)_{\beta,\rho}=\sum_{l=0}^\infty\frac{d_l
}{\Sigma_n}\frac1{r^{n-1}}\int\!\sigma_{\lambda
l}\frac{\lambda}{\e^{\beta \lambda}-1}f_{\lambda
l}^2(r^*)\dd\lambda \, ,
\end{equation}
where $d_l= \frac{(2l+n-2)(l+n-3)!}{l!(n-2)!}$ is the multiplicity
of the eigenvalue $l(l+n-2)$ of the squared total angular
momentum.

We expect that $\langle
T^\mu_{\phantom\mu\nu}\rangle_{\beta,\rho}$ has a structure
analogous to the one arising in the two dimensional case: a volume
term with singular behaviour for all values of $\beta$ except
$\beta=\beta_H$ and a boundary term, which appears when
considering a shell attached to the boundary. When performing an
analysis similar to the one of Section 4 we expect analogous
diverging contribution from the bulk and a potentially finite (see
discussion in Section 4) horizon contribution. Again, the leading
divergence could be traced back to the third principle. These
issues need a further investigation.

\section{Outlook and conclusions}

In this work we analysed the behavior of a quantum field at finite
temperature $T$ in the backgroud of a classical black hole. We
applied the brick wall approach of 't Hooft, introducing in
addition a parameter $\eta$, which fixes the boundary conditions
for the field on the brick wall and which can be interpreted as a
parametrization of the interaction of the field with gravity close
and behind the horizon. Focusing mainly on a 1+1 dimensional black
hole space-time, we computed both the energy and the entropy
densities. The energy density contains an $\eta$-dependent
boundary term, which is localized on the horizon and respects the
conservation and the trace anomaly of the energy-momentum tensor.
Taking into account the Boulware vacuum polarization, we have
shown that the energy density remains finite in the limit of
vanishing brick wall thickness, provided that $T$ equals the
Hawking temperature $T_H=\frac{\kappa_0}{2\pi}$. We recall in this
respect that in the original brick wall model the determination of
the brick wall thickness is made by requiring that the full
classical black hole entropy is due to the thermal atmosphere of
all the fundamental fields at the Hawking temperature\footnote{For
specific issues concerning the two dimensional case we refer to
\cite{Mann:1990fk}.}, which enters the model as an external input.

The entropy density shows analogous regular behavior for vanishing
brick wall thickness, provided we release the third principle of
thermodynamics in deriving the thermal entropy from the thermal
energy. We note here that there are several examples of incomplete
models that do not verify Nernst theorem. Likely the most known
and simple one is the perfect Boltzmann gas, where the entropy
diverges when $T\to 0$. In this case the problem is solved
observing that for sufficiently low temperatures the particle
interactions and quantum statistical effects cannot be neglected
and the model is no longer valid. This means that the physical
degrees of freedom of the system are not correctly identified by
the model - a situation which looks similar to ours.

The entropy we obtained has two main features. First, it is also
endowed with a boundary term localized on the horizon and
determined by the boundary condition. Second, the entropy in the
bulk is determined up to a function, since a new input,
substituting the third principle, is needed for its complete
determination. We believe that this input should come from a
better understanding of the interplay between quantum mechanics
and general relativity in systems containing horizons and black
holes.

For example, in the context of string theory, a recently quite
popular proposal \cite{Mathur:2005zp} suggests that the structure
of space-time inside the horizon is deeply quantum mechanical. In
this picture a model in which the semiclassical picture is
confined outside the horizon can be consistent. The specific
boundary condition can thus mimic some sort of interaction with
the physics inside the horizon. However, it could appear non
reasonable to perform the limit of vanishing brick wall thickness
since we cannot expect the transition between quantum and
semiclassical behaviour to happen sharply at the horizon. Also in
this picture some of our results are relevant: in particular the
term that gives rise to the ``would be'' boundary contribution,
still remain as a an indication of a peculiar accumulation of
degrees of freedom in the near horizon region.

\section*{Acknowledgements}
We would like to thank S. Liberati and V. Moretti for
encouragement and useful discussions. This research is supported by the
Italian MIUR under the program, ``Teoria dei Campi, Superstringhe
e Gravit\`a''.

\end{document}